
\magnification=1200
\hyphenpenalty=2000
\tolerance=10000
\hsize 14.5truecm
\hoffset 1.truecm
\openup 5pt
\baselineskip=24truept
\font\titl=cmbx12
\font\pic=cmr7

\def\({\left(}
\def\){\right)}

\font\mpic=cmmi7
\font\mmpic=cmmi5
\font\spic=cmsy7
\def\coolp{{\hbox{\mmpic cool}}}
\def\colp{{\hbox{\mmpic col}}}
\def\minpi{{\hbox{\mmpic min}}}
\def\dtpic{\hbox{\mpic\char'001 t}}
\def\simpic{\hbox{\spic\char'030}}
\def\ggpic{\hbox{\spic\char'035}}
\def\egpic{\hbox{\spic\char'025}}

\def\ref{\par\noindent\hangindent 20pt}

\def\mincir{\raise -2.truept\hbox{\rlap{\hbox{$\sim$}}\raise5.truept
\hbox{$<$}\ }}
\def\magcir{\raise -4.truept\hbox{\rlap{\hbox{$\sim$}}\raise5.truept
\hbox{$>$}\ }}
\def\rho{\varrho}
\def\Mdot{\dot M}

\def\MS{M_\odot/yr}

\def\Menv{M_{env}}
\def\etal{{\it et al.\/} }
\def\Tmax{T_{col}^{max}}
\def\Tmin{T_{col}^{min}}
\def\dt{\Delta t}
\null
\vskip 1.2truecm

\centerline{\titl ACCRETION RATES IN X--RAY BURSTING SOURCES}
\vskip 1.5truecm
\centerline{Iosif Lapidus $^1$\footnote{$^{\S}$}{The Royal Astronomical
Society Sir Norman Lockyer Fellow}, Luciano Nobili $^2$ and Roberto Turolla
$^2$}
\bigskip
\centerline{$^1$ Institute of Astronomy, University of Cambridge}
\centerline{Madingley Road, Cambridge CB3 0HA, UK}
\medskip
\centerline{$^2$ Department of Physics, University of Padova}
\centerline{Via Marzolo 8, 35131 Padova, Italy}

\beginsection ABSTRACT

We present estimates for the accretion rates in 13 X--ray bursting
sources which exhibit photospheric expansion, basing on theoretical models of
stationary, radiatively driven winds from neutron stars. The relatively high
values
obtained, $\Mdot_{acc}\magcir 10^{-9} \MS$, are in accordance with
theoretical limits for unstable helium  burning, and, at the same time,
almost never exceed the ``dynamical'' limit for stationary accretion,
$\sim 10 \Mdot_{Edd}$. The only exceptions are 1820-30, already known to
be a very peculiar object, and 1608-522; there are indications, however, that
in both sources, accretion could be non--stationary.

\bigskip\bigskip
\noindent
{\it Subject headings:\/} accretion, accretion disks \ -- \ stars: binaries
\ -- \ stars: individual \ -- \ stars: neutron \ -- \ X--rays: bursts
\bigskip
\centerline{Accepted for publication in the Astrophysical Journal}

\beginsection I. INTRODUCTION

Low--Mass X--ray Binaries (LMXBs) are widely believed to be powered by
the accretion of gas lost by the secondary, main sequence component of the
binary system onto a neutron star. This appears to be now a well established
point on both observational and theoretical grounds, although the observed
X--ray properties of
LMXBs can differ substantially from source to source, according to the
different physical conditions in the system. The presence of a
high neutron star magnetic field can produce a regular modulation
in the X--ray emission, as in X--ray pulsators, while other subclasses of
sources, in the large family of LMXBs, may be characterized by different
accretion scenarios. Within this picture, one of the most important parameters
in governing the overall appearance of the source, is certainly the mass
transfer rate, $\Mdot_{acc}$, from the secondary onto the neutron star.
It has been recently suggested (Kylafis \& Xilouris 1993)
that Super Soft Sources ultraviolet emission is produced in the very
dense, optically thick envelope which shrouds a neutron star accreting close
to the maximum allowed rate.
In this respect, X--ray bursters, too, should be characterized by quite
high values of the accretion rate  since unstable nuclear burning, which is
responsible for the bursting phenomenon, can take place only if
$\Mdot_{acc}$ exceeds some definite limit (Fushiki \& Lamb 1987; Taam
\etal 1993). The determination of accretion rates in X--ray bursting sources
with photospheric expansion, based on independent arguments,
would be quite useful both for testing the consistency of the helium burning
scenario, and for shedding light on various aspects of
evolution of close binaries. Any direct measurement of mass transfer rates
in LMXBs appears to be still beyond the capabilities of present
instrumentation and, up to now, the only, indirect, estimates come from the
comparison of the binary period variations with existing evolutionary
theories (see, for instance, Rappaport \etal 1987). In the present paper we
derive
accretion rates in LMXBs which show strong X--ray bursts with photospheric
expansion. During the expansion/contraction phase, in fact, a
supersonic wind is believed to be present and the envelope mass at the
beginning of the wind phase can be obtained by confronting the
observed spectral parameters with a set of theoretical wind models
(Nobili, Turolla \& Lapidus 1993); the knowledge of the interburst
time will then provide the accretion rate.
The analysis of the 13 sources for which sufficient data are available,
produced $\Mdot_{acc}\magcir 10^{-9}\MS$, in agreement with the current
idea that strong X--ray bursters should be characterized by high values
of the mass transfer rate in comparison with other LMXBs.

\beginsection II. PHOTOSPHERIC EXPANSION IN X--RAY BURSTERS

The most powerful observed X--ray bursts show a characteristic
temporal behavior. Usually the event starts with a sudden increase of
the X--ray intensity, with rise times less than  one second, followed by a
decrease, with a total duration of a few seconds.
After the precursor, a noticeable decay of the flux is observed, lasting up
to $\sim 10$ s in the strongest bursts, at the
end of which the luminosity can be as low as the persistent one.
The main part of the burst then begins. The increase of the X--ray intensity
first appears in the soft energy channels, and gradually becomes visible
in the harder bands. The color temperature increases while the X--ray
luminosity stays nearly constant at its maximum value, commonly associated with
the Eddington limit. A direct use of the relation $4\pi\sigma R_{col}^2
T_{col}^{4} = L\sim L_{Edd}$ shows that the typical size of the emitting
region,
$R_{col}$, after having reached a maximum, decreases in this phase until
$R_{col}$ is, again, comparable with the neutron stars radius $R_*$.
After the blackbody temperature has reached a maximum
(often above $\sim 2 $ keV), the decay starts with the progressive
decrease of the X--ray flux accompanied by a softening of the
spectrum, while the ``color'' radius remains approximately constant.
The last phase is quite similar to that one observed in weaker, type I, X--ray
bursts, where the energy released is below the Eddington limit during all the
burst and no expansion is observed.

Photospheric expansion during strong bursts is widely believed to be produced
by a supersonic outflow, driven by the large, super--Eddington luminosity
released by thermonuclear He burning at the base of the envelope.
In a very recent paper Nobili, Turolla \& Lapidus (1993, hereafter paper I,
see also Lewin, van Paradijs \& Taam 1993, hereafter LVPT, for a review of
earlier studies
on this subject) have presented a more complete model for radiative wind
acceleration during strong, type I, X--ray bursts which accounts properly
for both energy production by 3--$\alpha$ helium burning and Comptonization
heating--cooling in the outflowing envelope. One of the major results obtained
in that investigation was the discovery of a lower limit for the mass loss
rate $\Mdot_{min}$, below which no stationary, supersonic wind can exist. The
existence of
such a bound is due to a sort of ``preheating'' effect and could be of great
significance for the determination of some important physical parameters
in strong X--ray bursting sources. In paper I, in fact, we suggested that the
quasi--stationary expansion/contraction phase, during which $L\sim L_{Edd}$,
could be
thought as a sequence of steady wind models with decreasing $\Mdot$ which
terminates when $\Mdot_{min}$ is
reached. By taking this into account and if the maximum and minimum color
temperatures in the expansion phase, $\Tmax$, $\Tmin$,
are known from observational data, the initial envelope mass $\Menv$ can be
derived (see paper I for details); the mass accretion rate $\Mdot_{acc}$
follows once the time interval $\dt$ between two successive bursts with
photospheric expansion is available. The fitting of the theoretical $R_{col}$
--$T_{col}$ relation with the the observed one provides also an estimate of the
spectral hardening factor $\gamma$. In paper I this approach has been applied
to 4U/MXB 1820-30, mainly as a working example of the method.

Photospheric expansion during powerful bursts has been observed so far
in 14 sources of type I bursts, and in type II bursts from the Rapid
Burster 1730-335 (LVPT), although it can not be excluded that in some cases the
physical scenario may differ from that one outlined above.
Here we present results for all the bursters for which published data allow
a derivation of $\Tmax$, $\Tmin$ and $\dt$.

\beginsection III. INDIVIDUAL SOURCES

In this section we shall review the main observational properties of all
sources of type I bursts in which photospheric expansion has been
detected. Our primary goal
is to extract from observational data the maximum and minimum values of
the color temperature during the bursts with radius expansion together with
the interburst time. As we discussed in the previous section, in fact, if these
three parameters are known both the hardening factor and the accretion rate
onto the bursting neutron star can be derived from the sequence of stationary
wind models. The set of models we shall use below
refers to a neutron star of mass $M_* =1.5 M_\odot$ and radius $R_* =
13.5$ km. For 1820-30 there is a strong evidence that the secondary
is an evolved, helium--rich star, so helium wind models are used, while,
in comparing data relative to all other sources, ``solar'' composition models
seem more appropriate. For each source we derive $\gamma$, $\Mdot_{acc}$ and
compare the adiabatic cooling time of the envelope, $t_{cool}$, with the
interburst time. As it was discussed in paper I, accretion can be regarded as
stationary only if $t_{cool}\ll\dt$.

\medskip
\centerline{\it a) 0748-676}

During 9 observations with EXOSAT a total of 37 bursts were detected
(Gottwald \etal 1986).
Bursts with expansion always had $\dt > 5$ hr, and they occurred only when
the persistent flux was high, $\sim 2\times 10^{-9}$ erg$\, {\rm cm}^{-2}
{\rm s}^{-1}$.
In the paper by Gottwald \etal the data of all bursts are mixed
together, and only average figures of interest for us may be
extracted: $T_{col}^{max}\approx 2.2$ keV, and $T_{col}^{min}\approx 1.4$
keV.  Then $\gamma=1.68$ and
$\Menv=4\times 10^{22}$ g, with an accretion rate of
$\Mdot_{acc}\approx 3.5\times 10^{-8} (5 hr/\dt)\MS$.
The usual assumption that the
persistent flux is due to the conversion of gravitational potential energy
appears to be justified because the cooling time of the
envelope is $t_{cool}\sim 4\times 10^3\,{\rm s}
\,< \dt$ and, therefore, accretion
is stationary, contrary to what occurs in 1820-30 (see paper I and section IV).

Gottwald \etal (1987) observed also the source in a low state when no expansion
occurred
during bursts. These observations seem to confirm the suggestion by Fujimoto
\etal (1987) that incomplete nuclear burning is responsible for bursts
with short $\dt$, while in bursts with expansion (which show larger $\dt$)
nearly all the fuel is burned out.

\medskip
\centerline {\it b) 1516-569/Cir X--1}

EXOSAT observations (Tennant, Fabian, \& Shafer 1986)
revealed just one burst with photospheric expansion. The spectral data give
$T_{col}^{max}\approx 2.2$ keV, $T_{col}^{min}\approx 1.5$ keV, which implies
$\gamma=1.68$, $\Menv\approx 3\times
10^{22}$ g.  The time interval separating this burst from the previous one
is unknown, but it can be estimated
$\sim 4$ days. The estimated accretion rate $\Mdot_{acc}\approx 1.4\times
10^{-9}
\, \MS$ is below the critical value and accretion should be stationary since
$t_{cool}\simeq 2.5\times 10^3\,{\rm s} \ll \dt$.

\medskip
\centerline {\it c) 1608-522}

The observational data of Hakucho (Murakami \etal 1987) contain the record
of a burst with radial expansion that occurred on April 8, 1980. The bursts
frequency in April--May was $\nu_{b}=0.09\pm 0.03\, {\rm hr}^{-1}$.
The
blackbody fitting of spectra gives $T_{col}^{max}\approx 3.5-4.0$ keV,
$T_{col}^{min}\approx 1.2$ keV. The hardening factor is $\gamma=2.67$,
and a quite large value of the envelope mass is obtained, $\Menv =2\times
10^{23}$ g. The high cooling time of such an envelope, $t_{cool}\sim 10^5
{\rm s}\,\magcir \dt$,
and the high, largely super--Eddington accretion rate
$\Mdot_{acc}\approx 8\times 10^{-8} \MS$ make the source, at that period of
time, more similar to 1820-30 than to the ``normal'' bursters with
$\Mdot_{acc}\mincir\Mdot_{Edd}$.

Later observations by Tenma in 1984 indicate a decrease of the
accretion rate in this system. Nakamura \etal (1989) reported a
series of bursts, two of which (G and J in their notation)
showed radial expansion. In burst G the fitting temperatures
were $T_{col}^{max}=2.5$ keV, $T_{col}^{min}=1.8$ keV, resulting in
$\gamma=1.91$ and $\Menv=2.5\times  10^{22}$ g.
In the period of time preceeding burst G, the average
accretion rate was $\Mdot_{acc}\simeq 6\times 10^{-9} \MS$. Burst J
was stronger, with color temperatures  $T_{col}^{max}=2.9$ keV and
$T_{col}^{min}=1.5$ keV implying $\gamma=2.21$,
$\Menv=8\times 10^{22}$ g and a super--Eddington
accretion rate $\Mdot_{acc}\simeq 1.7\times 10^{-8} \MS$.
The accretion regime was, probably, stationary in both bursts since
$t_{cool}\sim 1.7\times 10^3\,{\rm s}\,\ll \dt = 6.7\times 10^4$ s (burst G),
and $t_{cool}\sim 2\times 10^4\,{\rm s}\, \mincir \dt = 7.4\times 10^4$ s
(burst J).

Photospheric second--range oscillations, observed
in 1608-522 during the long ($\sim 12$ s) flat top of the
light curve, are discussed in a separate paper (Lapidus, Nobili \& Turolla
1994).

\medskip

\centerline{\it d) 1636-536}

This object is one of most ``reliable'' sources of bursts and has been
extensively observed by Hakucho, Tenma and EXOSAT. As far as Tenma
observations are concerned, there is only one burst, denoted as burst H by
Inoue \etal (1984), which data can enable us to
determine $\Menv$ and $\Mdot_{acc}$. In all the other 11 bursts they
reported, either there is no expansion or
(for bursts D and E) the value $\Delta t$
before the burst with expansion
is unknown since observations were not continuous.
Burst H had $T_{col}^{max}\sim 2.9$ keV, $T_{col}^{min}\sim 2$ keV
which correspond to $M_{env}\sim 3.2\times 10^{22}$ g. With an interburst
time $\Delta t \sim 22$ hr, we obtain
$\Mdot_{acc}\sim 6.4\times 10^{-9}\, {M_{\odot}/yr}$.

An extensive analysis of numerous bursts from
1636-536 was performed by Lewin \etal (1987), using EXOSAT data.
Expansion was found in 3 bursts only, those ones with numbers 11, 26 and
27. Unfortunately no blackbody fitting during the expansion/contraction phase
was performed for these bursts and the re-examination of archive data
may provide valuable informations for this source.
\medskip

\centerline {\it e) 1715-321}

Tenma observations (Tawara, Hayakawa \& Kii 1984)
have registered only one burst with expansion, and $\Delta t$ is thus
unknown. It may be that there were no bursts in the 3 months preceeding the
burst under consideration, but it is also equally possible that
$\Delta t \sim 10$ days, as it has been in 1979.
The burst had $T_{col}^{max}\sim 1.1$ keV, $T_{col}^{min}\sim 0.25$ keV,
corresponding to $\gamma = 0.8$ and $M_{env}\sim 5\times 10^{23}$ g. With
$\Delta t \sim 10$ d, we obtain
$\Mdot_{acc}\sim 9.2\times 10^{-9}\, \MS$
while for $\Delta t
\sim 3$ months it is $\Mdot_{acc}\sim 10^{-9} \, \MS$.
The cooling time for
such an envelope is $t_{cool} \sim 10$ days, so the situation
may be fairly stationary for both values $\Delta t$ considered.

On July 20, 1982, Hakucho observed a very long ($\sim 300$ s) burst
with precursor, in which the luminosity stayed  at its maximum value for almost
100 s (Tawara \etal 1984a). The blackbody fitting gave
$T_{col}^{max}=3.0\pm 0.2$ keV, $T_{col}^{min}=1.1\pm 0.1$ keV. After this
burst, in 29 days of observations, no events were detected (although the
observations were not continuous, the effective
observation time was $\sim 120$ hr). Generally saying, this source
shows a weak activity, having produced only 3 bursts in 20 days in
1979 and a total of 4 bursts in 155 days during 1979--82. Therefore,
although the precise value of $\dt$ for the burst we consider is
unknown, a value $\sim 10$ d seems to be appropriate. We obtain
$\gamma=2.29$, $\Menv \approx 2\times 10^{23}$ g and the accretion rate
before the burst was then $\Mdot_{accr} \approx 3.7\times 10^{-9}\,(10d/\dt)
\MS$, of the same order as with Tenma data.

\medskip

\centerline{\it f) 1724-307/Terzan 2}

The only burst with expansion from this source was, probably, registered by
Grindlay \etal (1980) during a 30 minute Einstein observation.
The blackbody fitting of spectra produced $T_{col}^{max}\simeq 3.2$ keV,
$T_{col}^{min}\simeq 2$ keV, resulting in the hardening factor
$\gamma=2.44$  and $\Menv\sim 5\times 10^{22}$ g. The interburst
time preceeding this burst is unknown, but after this event  EXOSAT did
not register any bursts in 12 hr of continuous monitoring.
The value of accretion rate, $\Mdot_{accr}\sim
10^{-8} \times (1d/\dt)\MS $ may be equally sub-- or super--Eddington,
and no conclusion can be reached at the present stage.

\medskip

\centerline {\it g) 1728-337}

The source was extensively observed by SAS--3 in 1976--78, and a total of 60
bursts were registered (Hoffman \etal 1980).
The average $\dt$ was $3\div 8$ hr.
In only one of the bursts (on June 8, 1977) the expansion occurred.
Before this event there were no bursts in a 56 hr period, although some
bursts may have been missed because the Earth occulted the source roughly
one--third of the time. The spectral fitting parameters were
$T_{col}^{max}=2.6$ keV, $T_{col}^{min}=1.7$ keV which give
$\gamma=1.98$, $\Menv\sim 4\times 10^{22}$ g and,
assuming the 56 hr value as the actual $\dt$, the
accretion rate is
$\Mdot_{acc}\sim 3\times 10^{-9} (56hr/\dt)\, \MS$, just below the
Eddington level. The thermal regime of the envelope between bursts is
completely stationary since $t_{cool}\sim 1.2$ hr.

\medskip
\centerline{\it h) 1743-29}

The source is one of the Galactic Center X--ray bursters and it was observed
with SAS--3. All the bursts reported by Lewin \etal (1976) had double and
triple peaks. The average $\dt$ was $\sim 1.5$ d. Unfortunately, the spectral
fitting was impossible due to the poor quality of the data.

\medskip
\centerline{\it i) 1746-370/NGC 6441}

Sztajno \etal (1987) detected 2 bursts with expansion, separated by
8.5 hr, during a continuous 12 hr EXOSAT observation. The spectrum fitting
parameters of both bursts are quite similar, $T_{col}^{max}=2.2$ keV,
$T_{col}^{min}=1.2$ keV. The application of our theoretical model yields
$\gamma=1.68$,
$\Menv\simeq 6.5\times 10^{22}$ g, and an  accretion rate $\Mdot_{acc}
\simeq 3.4\times 10^{-8} \MS$. The high accretion rate is
in agreement with a cooling time scale
$\sim 10^4$ s $\sim \dt$.

The characteristics of the strongest bursts in 1746-370 seem to be not
changing in time. In fact, a
re-analysis of 2 bursts observed earlier with SAS--3 (Li \& Clark 1977) gives
$T_{col}^{max}= 2.2$, 2.4 keV, values quite similar to those derived
from EXOSAT data.

\medskip
\centerline{\it j) 1812-12}

The observations with Hakucho (Murakami \etal 1983) revealed two
bursts with photospheric expansion, on 1982, August 18 and 22.
The spectrum
fitting produced $T_{col}^{max}\sim 2.5$ keV; the data about the earlier
phases of the burst are far poorer and do not
allow the determination of $T_{col}^{min}$. Assuming, nevertheless, a
value of $T_{col}^{min} \sim 1.2 $ keV as characteristic from other
similar sources, we obtain $\gamma=1.9$, $\Menv \sim
9\times 10^{22}$ g and $\Mdot_{acc} \sim 3.3\times 10^{-9} (5d/\dt)\MS$.

\medskip
\centerline{\it k) 1820-30/NGC 6624}

The source is located in NGC 6624, and this provides a reliable
estimate for its distance of $6.4 \pm 0.6$ kpc (Vacca, Lewin \& van Paradijs
1986).
It exhibits a 685-seconds periodicity, first discovered by
Stella \etal (1987) with EXOSAT, and interpreted as an evidence for
orbital motion.
The orbital period is the shortest one known in LMXBs (see
Parmar \& White 1988 for a review) and is consistent with a scenario in which
the secondary is a low--mass, helium--rich degenerate star (Rappaport \etal
1987). Data on the time evolution of this
period are somewhat controversial. It was observed to decrease over the
years 1976--1991 (Sansom \etal 1989; Tan \etal 1991; van der Klis \etal
1993a) with a time scale of $\sim 10^{7}$ yr, which might be caused by
gravitational acceleration in the cluster potential or by a distant third
companion in a hierarchical triple. The standard scenario, involving mass
transfer from a Roche lobe--filling degenerate dwarf in an 11--minutes orbit
around a neutron star, predicts a secular increase of the orbital period
$> 8.8\times10^{-8}\, {\rm yr}^{-1}$ (Verbunt 1987; Rappaport \etal 1987)
rather then a decrease. The most recent ROSAT observations (van der Klis
\etal 1993b) do not provide any evidence of significant
period decrease.
It has been proposed that the secular
variation observed in 1976--1991 could have been dominated by
some changes in the position and shape of an occulting bulge on the disk rim.
The phase shifts caused by this mechanism could, in principle, mimic the
orbital period decrease, while the real period has been indeed increasing,
according to the standard theory.

As far as the evidence of photospheric expansion in this source is
concerned, there are two series of observations which can
provide information about the maximum and minimum color temperatures reached
in the expansion phase, and the interburst time.
The first is the series of
bursts observed by Clark \etal (1977) with SAS--3 in May 1975 and March
1976 and the second is the sequence of 7 successive bursts with photospheric
expansion observed by Haberl \etal (1987) with EXOSAT. The analysis of the
latter data was presented in paper I and provided the unusually high values of
$M_{env} \sim 9\times 10^{23}$ g and $\Mdot_{acc}
\sim 10^{-6} M_{\odot}$/yr.

Vacca, Lewin \& van Paradijs (1986) reported 6 bursts with photospheric
expansion in SAS--3 observations.  The satellite had two
independent detectors. In some bursts both of them were providing enough
information for the blackbody fitting, while, in other cases, only the
data from one detector were of satisfactory quality. The estimates of
the accretion rate are summarized in Table~1. The notation
N/A stands for cases when $T_{col}$ was not measured; the bursts are numbered
as in  Clark \etal (1977).
In spite of being somewhat lower than the figure obtained from EXOSAT data,
the present values $\Mdot_{acc}$ are always in excess of the Eddington value
for
He--rich matter, $\Mdot_{Edd}\sim 7\times 10^{-9}\, {\rm M_{\odot}/yr}$.
This confirms the suspicion that MXB 1820-30 is a very
peculiar source which
shows the highest mass transfer rate between X--ray bursters. Since bursters
distinguish themselves among LMXBs because of high accretion rates, MXB
1820-30 is, definitely,  one of the observed sources with highest
accretion rates. Together with the
still unexplained abnormal behavior of the orbital period,
our results make this source even more attractive for further
investigations.

\medskip
\centerline{\it l) 1850-087/NGC 6712}

The observations with SAS--3 revealed three bursts from the source, the
first two separated by 17 hr and without signs of expansion, and the third,
strongest one, with  expansion (Hoffman \etal 1980). The precise value
of $\dt$ for the third burst is unknown, but a value $\sim 1$ d seems
to be appropriate.
The spectral fitting produced $T_{col}^{max}\simeq 2.7$ keV,
$T_{col}^{min}\simeq 1.6 $ keV. Then the hardening factor is $\gamma=2.1$,
and $\Menv=5\times 10^{22}$ g. The accretion rate is
$\Mdot_{accr}\simeq 9\times 10^{-9}(1d/\dt) \MS$. Being close to
the Eddington limit for  He--rich matter, and only twice that one for a pure
hydrogen envelope, such an accretion rate seems to be consistent with
the fairly low persistent flux.
\vfill\eject
\centerline{\it m) 1905+000}

Chevalier \& Ilovaisky (1990) made a detailed spectral analysis of one
burst with photospheric expansion detected during a continuous 19 hr EXOSAT
observation. Although being non--perfect, the blackbody fitting gave
$T_{col}^{max}=3.5\pm 0.7$ keV, $T_{col}^{min}=1.7$ keV, resulting in
$\gamma=2.67\pm 0.53$ and $\Mdot_{acc} \sim 10^{-8} (1d/\dt) \MS$.
The thermal regime of the envelope should be stationary since
$t_{cool}\sim 10^4$ s.

\centerline{\it n) 2127+119/M15}

One very strong burst with expansion was reported by Dotani \etal (1990) and
Van Paradijs \etal (1990). Similarly to what was observed in 1608-522, a
series of photospheric oscillations were detected with Ginga during the
first $\sim 30$ s of the burst in this source; this issue is addressed to
in Lapidus \etal (1994).
The blackbody spectral fit
provided $T_{col}^{max}\simeq 2.8$ keV,
$T_{col}^{min}\simeq 1.4$ keV, although, like in some other
sources, the spectra deviated from a blackbody in a way that
cannot be described in terms of a spectral hardening alone (see discussion
below).
Applying our usual technique, we obtain $\gamma=2.1$
and $\Menv\simeq 8\times 10^{22}$ g. The expectation time of a burst is
unknown, but a value $\sim$ few days seems to be appropriate.
The estimated accretion rate $\Mdot_{accr}\simeq 5\times 10^{-9}(3d/\dt) \MS$
does not exceed, probably, the Eddington level.

\beginsection IV. DISCUSSION AND CONCLUSIONS

The results obtained in the previous section for 13 sources which exhibit
photospheric expansion are summarized in Table 2. It can be seen from the table
that, in several cases, the derived values of the accretion rate exceed the
Eddington rate
$\Mdot_{Edd}\sim 4\times 10^{-9}\MS$ for a ``solar'' composition. Such a result
not necessarily contradicts standard steady accretion theories since, assuming
spherical accretion, the total power released is $\sim (GM_*/R_*)
\Mdot_{acc}\sim \Mdot_{acc}c^2/6$ for a neutron star radius
nearly three times the gravitational radius, as in our wind models.
This means that accretion rates
up to 6 times the critical one are still possible. This limit can be even
higher if, as it seems more plausible, an accretion disk is formed around the
neutron star. In disk accretion, in fact, only about half of the gravitational
energy is released in the vicinity of the star surface (the rest being
radiated away
in the extended disk), and it is only this part of the energy release which
can place a limit on the accretion rate. This argument would bring the upper
limit for the permitted value of $\Mdot_{acc}$ to $\sim 10 \Mdot_{Edd}$.
Larger accretion rates are still possible in non--stationary accretion, for
which the Eddington limit does not apply; in this case gravitational energy
is temporarily stored in the accreted material in the form of internal energy.

We note that, although our estimates of the accretion rate are model dependent,
they do not imply any precise accretion scenario, that is to say they hold
the same no matter how the gas is accreted onto the neutron star and where it
comes from. It is apparent from table 2 also that the main source of
uncertainty in the determination of $\Mdot_{acc}$ is $\dt$, which is often
poorly known, while the evaluation of both $\gamma$ and $\Menv$ appears to be
more reliable.
It seems reasonable to divide the sources in three classes,
according to the estimated value of $\Mdot_{acc}$: the range
$\Mdot_{acc}\mincir \Mdot_{Edd}$ defines group I, $\Mdot_{Edd}\mincir
\Mdot_{acc}\mincir 10\Mdot_{Edd}$
group II and $\Mdot_{acc}\magcir 10\Mdot_{Edd}$ group III. Group I contains 4
sources: 1516-569, 1715-321, 1728-337 and 1812-12;
in group II there are 8 sources: 0748-676, 1608-522 (as observed with Tenma),
1636-536, 1724-307, 1746-370 (marginally belonging to group III), 1850-87,
1905+000, 2127+119.
Out of the 13 sources listed in table 2, only 1820-30 definitely belongs to
group III, confirming its peculiar nature, together with 1608-522 (as observed
with Hakucho). In paper I it was suggested that the highly super--Eddington
accretion rate in 1820-30 could be reconciled with standard accretion scenarios
on the basis of the unstationary nature of the accretion process. The
comparison of the characteristic time needed to radiate away the gravitational
energy release, which increases with $\Menv$, with the interburst time has
shown, in fact, that, in this source, the envelope has no time to cool between
two successive bursts, and therefore the persistent flux could not be directly
related to $\Mdot_{acc}$. The same argument applies to 1608-522 in the
``high'' state, as observed by Hakucho, when $t_{cool}\sim 3 \dt$. It is
remarkable that in all cases where $\Mdot_{acc}$ exceeds $\sim 10
\Mdot_{Edd}$, it is $t_{cool}\magcir \dt$.
This can be regarded as a confirmation of
our model because the evaluation of the cooling time is completely independent
on $\dt$. A further evidence is provided by the data of 1746-370, which
seems to lie at the border of classes II and III and has
$\Mdot_{acc} \sim 10 \Mdot_{Edd}$, $t_{cool}\sim \dt$.
Theoretical limits on the accretion rate
for the onset of the helium thermonuclear runaway were placed by
Fushiki \& Lamb. According to their analysis of the stability of nuclear
burnings in a static, solar composition envelope, a helium flash can
be produced only if $10^{-10}\mincir\Mdot/(\MS)\mincir 10^{-8}$.
The values we have derived are indeed within this range, with the
exception of 1820-30 and, possibly, 1608-522,
and this seems to be consistent with a picture in which unstable He burning
is responsible for strong bursts with photospheric expansion. We
note that, in the case of 1820-30, the accreted material is
believed to be helium--rich, so that a straightforward application
of Fushiki \& Lamb's results to this source is not appropriate. As far as
1608-522 is concerned, the accretion rate turns out to be well above
$ 10^{-8}\MS$ only
in one event, while other two observations provide values in the above
range.
The present estimates
of accretion rates onto bursting neutron stars are also not in contradiction
with the  models of Taam \etal, who followed the time evolution of
recurrent hydrogen thermonuclear flashes. They found, in fact, that accretion
rates $\approx  10^{-10}\MS$ give rise to repeated H flashes while suggesting
that unstable He burning may be produced only for larger $\Mdot_{acc}$
because the high transfer rate can inhibit convective mixing and increase
the envelope cooling time.

A definite improvement on the present method for estimating the accretion rate
will come from self--consistent frequency--dependent transfer calculations
which are now under way. They will provide, in fact, the hardening ratio for
each wind model. This sequence of $\gamma$'s will substitute the
average constant value of $\gamma$ used so far, which was itself derived by
matching spectral data with the models. The comparison of the computed
hardening ratios
with the present ones would also provide a further test for the consistency
of our approach. We note, however, that the values of $\gamma$ listed in
table 2 are quite reasonable. As we discussed in paper I about 1820-30,
$\gamma\sim 2$ is in good accordance with recent transfer calculations in
expanding envelopes (Lapidus 1991) and, for all sources we examined, it is
$1.5\leq\gamma\leq 2.7$. The only exceptions are the Tenma observation of
1715-321, which give $\gamma = 0.8$, while
a ``normal'' value, $\sim 2.3$, was obtained from Hakucho's data.
Frequency--dependent calculations will also allow the comparison of the
self--consistently computed emerging spectrum with observations. This appears
to be of particular relevance for strong bursts in which a pure
blackbody fitting was often found to be unsatisfactory.
Frequency--dependent radiative transfer calculations in
expanding envelopes performed so far (Lapidus 1991; Titarchuk 1993; Turolla, \
Nobili, Zampieri \& Lapidus, in preparation), show, in fact, that a
definite soft excess in addition to the general spectral hardening should be
expected.

\beginsection ACKNOWLEDGEMENTS

We thank Philip Podsiadlowski for useful discussions.
One of us (I.~L.) gratefully acknowledges support from the Royal
Astronomical Society and Consiglio Nazionale delle Ricerche--Gruppo Nazionale
di Astronomia.
\vfill\eject

\beginsection REFERENCES

\ref{Clark, G.W., Li, F.K., Canizares, C., Hayakawa, S., Jernigan, G., \&
Lewin, W.H.G. 1977, MNRAS, 179, 651}
\ref{Chevalier, C., \& Ilovaisky, S.A. 1990, A\&A, 228, 115}
\ref{Dotani, T., Inoue, H., Murakami, T., Nagase, F., Tanaka, Y., Tsuru,
T., Makishima, K., Ohashi, T., \& Corbet, R.H.D. 1990, Nature, 347, 534}
\ref{Fujimoto, M.Y., Sztajno, M., Lewin, W.H.G., \& Van Paradijs, J. 1987,
ApJ, 319, 902}
\ref{Fushiki, I., \& Lamb, D.Q. 1987, ApJ, 323, L55}
\ref{Gottwald, M., Haberl, F., Parmar, A.N., \& White, N.E. 1986, ApJ, 308,
213}
\ref{Gottwald, M., Haberl, F., Parmar, A.N., \& White, N.E. 1987, ApJ, 323,
575}
\ref{Grindlay, J.E., Marshall, H.L., Hertz, P., Soltan, A., Weisskopf,
M.C., Elsner, R.F., Ghosh, P., Darbo, W., \& Sutherland, P.G. 1980, ApJ,
240, L121}
\ref{Haberl, F., Stella, L., White, N.E., Priedhorsky, W.C., \& Gottwald,
M. 1987, ApJ, 314, 266}
\ref{Hoffman, J.A., Cominsky, L., \& Lewin, W.H.G. 1980, ApJ, 240. L27}
\ref{Inoue, H., Waki, I., Koyama, K., Matsuoka, M., Ohashi, T., Tanaka, Y.,
\& Tsunemi, H. 1984, PASJ, 36, 831}
\ref{Kylafis, N.D. \& Xilouris, E.M. 1993, A\&A, 278, L43}
\ref{Lapidus, I.I. 1991, ApJ, 377, L93}
\ref{Lapidus, I., Nobili, L., \& Turolla, R. 1994, ApJ Lett., submitted}
\ref{Lewin, W.H.G., Li, F.K., Hoffman, J.A., Doty, J., Buff, J.,
Clark, G.W., \& Rappaport, S. 1976, MNRAS, 177, 93P}
\ref{Lewin, W.H.G., Penninx, W., Van Paradijs, J., Damen, E., Sztajno, M.,
Tr\"{u}mper, J., \& Van der Klis, M. 1987, ApJ, 319, 893}
\ref{Lewin, W.H.G., Van Paradijs, J., \& Taam, R.E. 1993, Space Sci.
Rev., 62, 223 (LVPT)}
\ref{Li, F.K., \& Clark, G.W. 1977, IAU Circular No. 3095}
\ref{
Murakami, T., Inoue, H., Koyama, K., Makishima, K., Matsuoka, M., Oda, M.,
Ogawara, Y., Ohashi, T., Makino, F., Shibazaki, N., Tanaka, Y., Hayakawa,
S., Kunieda, H., Masai, K., Nagase, F., Tawara, Y., Miyamoto, D.S.,
Tsunemi, H., Yamashita, K., \& Kondo, I. 1983, PASJ, 35, 531}
\ref{Murakami, T., Inoue, H., Makishima, K., \& Hoshi, R. 1987, PASJ, 39,
879}
\ref{Nakamura, N., Dotani, T., Inoue, H., Mitsuda, K., \& Tanaka, Y.
1989, PASJ, 41, 617}
\ref{Nobili, L., Turolla, R., \& Lapidus, I. 1993, ApJ, submitted}
\ref{Parmar, A.N. \& White, N.E. 1988, Mem.S.A.It., 59, 147}
\ref{Rappaport, S., Nelson, L., Joss, P., \& Ma, C.--P. 1987, ApJ, 322, 842}
\ref{Sansom, A., Watson, M., Makishima, K., \& Dotani, T. 1989, PASJ, 41, 591}
\ref{Sztajno, M., Fujimoto, M.Y., Van Paradijs, J., Vacca, W.D., Lewin,
W.H.G., Penninx, W., \& Tr\"{u}mper, J. 1987, MNRAS, 226, 39}
\ref{Stella, L., White, N.E., \& Priedhorsky, W. 1987, ApJ, 312, L17}
\ref{Taam, R.E., Woosley, S.E., Weaver, T.A., \& Lamb, D.Q. 1993, ApJ, 413,
324}
\ref{Tan, J., Morgan, E., Lewin, W., Penninx, W., van der Klis, M., van
Paradijs, J., Makishima, K., Inoue, H., Dotani, T., Mitsuda, K. 1991, ApJ,
374, 291}
\ref{Tawara, Y., Kii, T., Hayakawa, S., Kunieda, H., Masai, K., Nagase, F.,
Inoue, H., Koyama, K., Makino, F., Makishima, K., Matsuoka, M., Murakami,
T., Oda, M., Ogawara, Y., Ohashi, T., Shibazaki, N., Tanaka, Y., Miyamoto,
S., Tsunemi, H., Yamashita, K., \& Kondo, I. 1984a, ApJ, 276, L41}
\ref{Tawara, Y., Hayakawa, S., \& Kii, T. 1984b, PASJ, 36, 845}
\ref{Tennant, A.F., Fabian, A.C., \& Shafer, R.A. 1986, MNRAS, 221, 27P}
\ref{Titarchuk, L. 1993, ApJ, submitted}
\ref{Vacca, W.D., Lewin, W.H.G., \& van Paradijs, J. 1986, MNRAS, 220, 339}
\ref{Van der Klis, M., Hasinger, G., Dotani, T., Mitsuda, K., Verbunt, F.,
Murphy, B., van Paradijs, J., Belloni, T., Makishima, K.,
Morgan, E., \& Lewin, W. 1993a, MNRAS, 260, 686}
\ref{Van der Klis, M., Hasinger, G., Verbunt, F., van Paradijs, J.,
Belloni, T., \& Lewin, W.H.G. 1993b, A\&A, 279, L21}
\ref{Van Paradijs, J., Dotani, T., Tanaka, Y., \& Tsuru, T. 1990, PASJ, 42,
633}
\ref{Verbunt, F. 1987, ApJ, 312. L23}
\vfill\eject

%
\medskip
\baselineskip=24truept
\centerline{Table 1}
\smallskip
\centerline{Accretion Rates in 4U/MXB 1820-30 (SAS--3 data)}
\medskip\medskip
\hrule
$$\vbox{\tabskip=1em plus2em minus.5em
\halign to\hsize{#\hfil &\hfil # \hfil & \hfil # \hfil & \hfil # \hfil &
 \hfil # \hfil  & \hfil # \hfil & \hfil # \hfil & \hfil # \hfil\cr
& Burst $\#$ & & Detector~1& & Detector~2 & & \cr
& & & $\Mdot_{acc}\, ({M_{\odot}/yr})$ & & $\Mdot_{acc}\,
({M_{\odot}/yr})$ &
& \cr
%
%
\noalign{\smallskip}
 & 1  & & $1.7\times 10^{-7}$   &    &  $9.7\times 10^{-7}$ & & \cr
 & 4  & & N/A           &    &  $1.7\times 10^{-7}$ & & \cr
 & 6  & & $4.4\times 10^{-7}$   &    &  $1.2\times 10^{-6}$ & & \cr
 & 7  & & $2.8\times 10^{-7}$   &    &  N/A         & & \cr
 & 11 & & N/A           &    &  $1.3\times 10^{-7}$ & & \cr
 & 20 & & $1.4\times 10^{-7}$   &    &  $4.2\times 10^{-7}$ & & \cr
 }}$$
\hrule
\vfill\eject
%
\medskip
\centerline{Table 2}
\smallskip
\centerline{Accretion Rates and hardening ratios in X--ray bursting sources}
\medskip\medskip
\hrule
$$\vbox{\tabskip=1em plus2em minus.5em
\halign to\hsize{#\hfil &\hfil # \hfil & \hfil # \hfil & \hfil # \hfil &
 \hfil # \hfil  & \hfil # \hfil & \hfil # \hfil & \hfil # \hfil\cr
& Source & $\Mdot_{acc}$ & $\gamma$ & Notes & & & \cr
& & $(\MS )$ & & & & & \cr
%
%
\noalign{\smallskip}
 & 0748-676 & $\mincir 3\times 10^{-8}$ & 1.7 & {\pic only \ lower \ limit \
for \ \dtpic \ available}    &  & & \cr
 & 1516-569 & $\sim 10^{-9}$ & 1.7 & \dtpic \ {\pic only \ estimated} &  & &
\cr
 & 1608-522 & $8\times 10^{-8}$ & 2.7 & {\pic Hakucho, \ } $\hbox{\mpic t}
_{\coolp}\egpic \dtpic$
   &  & & \cr
 &          & $6\times 10^{-9}$  & 1.9 & {\pic Tenma \ burst \ G}& & &\cr
 &          & $2\times 10^{-8}$  & 2.2 & {\pic Tenma \ burst \ J}& & &\cr
 & 1636-536 & $6\times 10^{-9}$ & 2.2 & {\pic Tenma \ burst \ H}   &  & & \cr
 & 1715-321 & $4\times 10^{-9}$ & 2.3 & {\pic Hakucho}   &  & & \cr
 &          & $10^{-9}\div 10^{-8}$  & 0.8 & {\pic Tenma, \ \dtpic \ unknown, \
drastic \ spectral \ softening \ ?}& & &\cr
 & 1724-307 & $\sim 2\times 10^{-8}$ & 2.4 & \dtpic \ {\pic only \ estimated}
  &  & & \cr
 & 1728-337 & $\mincir 3\times 10^{-9}$ & 2.0 & {\pic only \ lower \ limit \
for \ \dtpic \ available}     &  & & \cr
 & 1746-370 & $3\times 10^{-8}$ & 1.7 & $\hbox{\mpic t}_{\coolp}\simpic
\dtpic$   &  & & \cr
 & 1812-12  & $\sim 3\times 10^{-9}$ & 1.9 & $\hbox{\mpic T}_{\colp}^{\minpi}$
\
{\pic unknown}
   &  & & \cr
 & 1820-30  & $ 10^{-6}$ & 1.5 & {\pic EXOSAT, \ He models, \ $\hbox{\mpic t}_
{\coolp}\ggpic\dtpic$}   &  & & \cr
 &   & $10^{-7}\div 10^{-6}$ & 2.3 & {\pic SAS--3 \
(see  \ table \ 1), \ He models}   &  & & \cr
 & 1850-87  & $9\times 10^{-9}$ & 2.1 &  \dtpic \ {\pic only \ estimated}  &
& & \cr
 & 1905+000 & $\sim 10^{-8}$ & 2.7 & \dtpic \ {\pic unknown}   &  & & \cr
 & 2127+119 & $\sim 5\times 10^{-9}$ & 2.1 & \dtpic \ {\pic unknown}   &  & &
\cr
 }}$$
\hrule
\vfill\eject
\bye